\documentstyle[prl,aps]{revtex}
\large
\begin{document}
\twocolumn
\draft
\title
{Gauge invariance and wave packet simulations\\
in the presence of dipole fields}
\author{Thierry Martin$^*$}

\address{CNLS, Theory Division, Los Alamos National Laboratory,
Los Alamos, NM 87545}
\maketitle


\begin{abstract}
A method for performing wave packet simulations in dipole fields
is presented. Starting from a Hamiltonian with non commuting terms,
a gauge transformation leads to a new Hamiltonian which
allows to calculate explicitly the evolution operator.
In this new gauge, the dipole field is fully included in the
{\it vector} potential. The method of Goldberg, Schwartz and
Schey based on the Caley form of the evolution operator
is then generalized, and the resulting scheme is applied to
describe a switching device. The device is composed of a well
region separated from a free region by a potential barrier,
such that one bound state and one quasi level are present.
For a particle initially in the ground state, transitions to
the quasi level are induced by applying a dipole field, and the
wave function can subsequently tunnel in the free region.
The probability to tunnel in the free region exhibits a plateaux
structure as the wave function is emitted by ``bursts'' after each
Rabi oscillation has been completed.
\end{abstract}
\smallskip
\bigskip
\pacs {}
\narrowtext
The pioneering work of Goldberg, Schey and Schwartz \cite{Goldberg}
on numerical simulations of wave packet evolution through
potential barriers and wells continues
to have a deep impact on the teaching community.
However, when this work first appeared, the technological
means to observe and characterize experimentally such phenomenon
were absent. One dimensional quantum mechanics was then
considered on a rather academic level. With the recent advances
in the fabrication of semiconductor nano structures,
one now has the means to design such structures
using molecular beam epitaxy techniques, etc... The motivation
for research in this field now comes from the hope that
quantum devices may one day be used as elementary building blocks
for the next generation of computers.
In the following, I will describe a method which allows to
simulate the evolution of a wave packet in a potential
landscape, in the presence of a dipole potential.
The method will be applied to describe a switching device
consisting
of a well, separated by a free region by a potential barrier.
Starting with a particle which is trapped in the well,
it is possible to induce a current in the free region
by appling a micro wave field which activates the particle to
a quasi level in the well. The tunneling current
can be tuned by adjusting the excitation frequency,
the amplitude of the external field
and the characteristics of the barrier.

While simulations of wave packet transmission through oscillating barriers
have been achieved in the past \cite{Haavig}, little is known for the more
realistic situation where the applied potential depends linearly on
the distance (electric field constant in space). This is relevant
when a micro wave field is applied to a semiconductor heterostructure.
 Our starting point is the Hamiltonian
$H=H_0 + \phi(x, t)$
with the free Hamiltonian:
\begin{equation}
H_0=-{\hbar^2\over 2m}{\partial^2\over\partial x^2}+V(x)
\label{stationary}
\end{equation}
and the dipole potential:
\begin{equation}
\phi(x, t)=\epsilon x\cos(\omega t)
\label{dipole}
\end{equation}
with $\epsilon$ the coupling strength and $\omega$ the driving frequency.
For simplicity, I will consider static potentials $V(x)$ which are
constant in a set of different regions separated by potential steps.
For more general situations, the potential $V(x)$ can be decomposed
as a series of infinitesimal steps \cite{Mendez}.
Given an initial state $\psi(x,0)$, the wave function at later times
is obtained by application of the evolution operator:
\begin{equation}
\psi(x, t)=U(t,0)\psi(x,0)
\end{equation}
{}From standard Quantum Mechanics textbooks \cite{Baym}, the evolution
operator is known to be:
\begin{equation}
U(t, t_0)=T\exp\biggl[-{i\over\hbar}\int_{t_0}^t  dt^\prime
(H_0+\phi(x, t^\prime))\biggr]
\label{evolution}
\end{equation}
where the symbol $T$ denotes the time ordering procedure.
In past calculations, the assumption that the time varying field $\phi(t)$
be a constant in a given region of space allowed to drop the time
ordering factor: $H_0$
commutes with $\phi(t)$ in this particular case.
For the Hamiltonian of Eqs. (\ref{stationary}) and (\ref{dipole}),
the time
ordering has to be kept because the kinetic energy does not
commute with the position operator $x$. How can
a numerical algorithm
similar to that of Goldberg, Schey and Schwartz (GSS) \cite{Goldberg}
be implemented? A ``brute force'' solution to
this problem is
to choose the discretization in time sufficiently small so that only
the lowest order terms ($O(H)$) in the exponential of Eq. (\ref{evolution})
are important: the time ordering does not affect the approximate
evolution operator for infinitesimal times:
\begin{equation}
U(t+ \delta t, t)\simeq 1-{i\over\hbar}\int_t^{t+\delta t}dt^\prime
(H_0+\phi(x, t))
\label{elementary}
\end{equation}
This ceases to be true when second order terms are taken into account.
In addition to this blemish, another problem arises:
in situations where a large spatial region is required,
such as in the case where a (small) barrier or well region
which specify the characteristic
energies of the problem, is coupled to a (large) continuum
region, the expectation
value $<x>$ of the position operator may take large values.
This puts further restrictions on the magnitude of the
time step for the simulation.

Nevertheless, a more elegant method, which exploits the gauge
invariance property of the
Hamiltonian, does not suffer from the same constraints.
The Hamiltonian of Eqs. (\ref{stationary}) and (\ref{dipole})
is written in a gauge where the vector potential $A=0$ and the
scalar potential is $\phi(x, t)$.
An alternative choice of gauge can be obtained as:
\begin{eqnarray}
A^\prime&=&A-{\partial\over\partial x}\chi\\
\phi^\prime&=&\phi+{1\over c}{\partial\over\partial t}\chi
\end{eqnarray}
Now, I choose the ``new'' gauge so that the scalar potential
$\phi^\prime=0$. This then yields:
\begin{eqnarray}
\chi(x, t)&=&-{c \epsilon x\over\omega}\sin(\omega t)\\
A^\prime(x, t)&=&{c \epsilon\over \omega}\sin(\omega t)
\end{eqnarray}
In this new gauge, the Hamiltonian is written as:
\begin{eqnarray}
H^\prime&=&{1\over 2m}\biggl({\hbar\over i}{\partial\over\partial x}
-{e\over c}A^\prime
\biggr)^2+V(x)\nonumber\\
&=&-{\hbar^2\over 2m}{\partial^2\over\partial x^2}+i{e\hbar\over mc}
A^\prime{\partial\over\partial x}+V(x)
+{e^2\over 2mc^2}A^{\prime 2}
\end{eqnarray}
The second line follows from the fact that $A^\prime$ commutes
with the momentum operator. In the new gauge, the wave function
$\psi^\prime(x, t)$ is related to the old one:
\begin{equation}
\psi^\prime(x, t)=\psi(x, t)e^{-ie\chi(x, t)/\hbar c}
\end{equation}
Note that the Hamiltonian $H^\prime$ does not suffer from the same
setbacks as $H$: in each constant region of $V(x)$ {\it all} terms
in the Hamiltonian $H^\prime$ commute with each other, and consequently, the
time ordering factor in the expression for the evolution operator
$U^\prime(t, t_0)$ can be dropped out:
\begin{eqnarray}
U^\prime(t, t_0)&=&\exp\biggl[-{i\over\hbar}\int_{t_0}^t dt^\prime
H^\prime(t^\prime)\biggr]\\
&=&\exp\biggl[-i{t-t_0\over\hbar}H_0-{g(t, t_0)\over m}
{\partial\over\partial x}-i\theta(t, t_0)\biggr]
\nonumber\end{eqnarray}
with
\begin{eqnarray}
g(t, t_0)&=&{\epsilon\over \omega^2}(\cos(\omega t)-\cos(\omega t_0))\\
\theta(t, t_0)&=&{e^2\over 2\hbar mc^2}\int_{t_0}^tdt^\prime
A^{\prime 2}(t^\prime)
\end{eqnarray}
Note that the last term in the exponential contributes only
a time  dependent phase factor to the evolution.
Therefore, $\theta(t, t_0)$ will be omitted in the following.

I now generalize the numerical scheme of GGS to the
case where a vector potential is present. The elementary
time step evolution for the wave function
$\psi^\prime$ is taken in the so-called Caley form:
\begin{equation}
\psi^\prime(t+\delta t) ={1-{i\over\hbar} \delta t H^\prime
\over 1+{i\over\hbar} \delta t H^\prime}
\psi^\prime(t)
\end{equation}
or, alternatively, in the ``old'' gauge:
\begin{equation}
\psi(t+\delta t) =e^{ie\chi(t+\delta t)/\hbar c}{1-{i\over\hbar}
\delta t H^\prime\over 1+{i\over\hbar} \delta t H^\prime}
e^{-ie\chi(t)/\hbar c}\psi(t)
\end{equation}
which has the advantage of being exact to order $(\delta t)^2$ and unitary.
Taking the convention $\hbar=1$, $m=1/2$, and choosing the discretized
variables $x=j \delta x~(j=0,1,...J)$, $t=n \delta t~(n=1,2...)$,
($\psi(x, t)\equiv \psi_j^n$), ($g(t)=\delta t g_n$), ($V(x)=V_j$),
this is rewritten
as:
\begin{eqnarray}
&&\psi_j^{n+1}(1+i{\delta t\over 2 } V_j)+i\lambda^{-1}(-\psi_{j+1}^{n+1}
+2\psi_j^{n+1}-\psi_{j-1}^{n+1})\nonumber\\
&&+i{\delta t\over 2 \delta x}(-ig_n)(\psi_{j+1}^{n+1}
-\psi_{j-1}^{n+1})
=\nonumber\\
&&+\psi_j^n(1-i{\delta t\over 2 } V_j)
-i\lambda^{-1}(-\psi_{j+1}^n
+2\psi_j^n-\psi_{j-1}^n)\nonumber\\
&&-i{\delta t\over 2 \delta x}(-ig_n)(\psi_{j+1}^n-\psi_{j-1}^n)
\label{scheme 1}
\end{eqnarray}
where $\lambda=2\delta x^2/\delta t$.
Introducing the quantity:
\begin{eqnarray}
\Omega_j^n&=&\psi_{j+1}^n(-1-i\delta x g_n)+\psi_j^n(2+\delta x^2V_j
+i\lambda)\nonumber\\
&~&+\psi_{j-1}^n(-1+i\delta x g_n)
\label{scheme 2}\end{eqnarray}
Eq. (\ref{scheme 1}) then takes the form:
\begin{eqnarray}
&&\psi_{j+1}^{n+1}(1+i\delta x g_n)+\psi_j^{n+1}(-2-\delta x^2V_j+i\lambda)
\nonumber\\
&&+\psi_{j-1}^{n+1}(1-i\delta x g_n)=\Omega_j^n
\end{eqnarray}
The above equation can be solved with the ansatz \cite{Goldberg}:
\begin{equation}
\psi_{j+1}^{n+1}=e_j^n\psi_j^{n+1}+f_j^n
\label{ansatz}
\end{equation}
Which yields expressions for the quantities $e_j^n$ and $f_j^n$:
\begin{eqnarray}
e_j^n&=&-{1-i\delta x g_n\over 1+i\delta x g_n}{1\over e_{j-1}^n}+
{2+\delta x^2V_j-i\lambda\over 1+i\delta x g_n}\label{r e c}a\\
f_j^n&=&{1-i\delta x g_n\over 1+i\delta x g_n}{f_{j-1}^n\over e_{j-1}^n}
+{\Omega_j^n\over 1+i\delta x g_n}\ref{r e c}b
\end{eqnarray}
Boundary conditions for the above quantities are now needed. These are
obtained from the boundary condition on the wave function
($\psi_0^n=\psi_J^n=0$ for all $n$):
\begin{equation}
\psi_2^{n+1}={2+\delta x^2V_1-i\lambda\over 1+i\delta x g_n}\psi_1^{n+1}
+{\Omega_1^n\over 1+i\delta x g_n}
\end{equation}
which in turn implies
\begin{eqnarray}
e_1^n&=&{2+\delta x^2 V_1-i\lambda\over 1+i\delta x g_n}\\
f_1^n&=&{\Omega_1^n\over1+i\delta x g_n}
\end{eqnarray}
Using Eqs. (\ref{r e c}a--\ref{r e c}b), one can now obtain $e_j^n$ and $f_j^n$
for $j=2,..,J-1$.
Similarly, the boundary condition for $\psi_{J-1}^{n+1}$
yields:
\begin{equation}
\psi_{J-1}^{n+1}=-f_{J-1}^n/ e_{J-1}^n
\end{equation}
One can now update the wave function for $j=J-2,...,2$ using
the ansatz of Eq. (\ref{ansatz}):
\begin{equation}
\psi_j^{n+1}=(\psi_{j+1}^{n+1}-f_j^n)/e_j^n
\end{equation}
which completes the numerical scheme.

I now apply this method to a specific example. It has long been known
\cite{Landau} for a two level system that if a harmonic perturbation is
applied such that the excitation frequency matches approximately
the spacing between the two levels, the probability of occupation of a given
level undergoes
slow oscillations. Such oscillations are known in optics
as Rabi oscillations.
If the system is started in the ground state, and the
dipole potential is switched on at $t=0$, the system undergoes
transitions to the excited state,
but will ultimately
return in the ground state after a time $T_R=2\pi\hbar/|F_{12}|$
(resonant case), where
$F_{12}$ is the matrix element of the harmonic perturbation.
These slow oscillations will be exploited
to construct a switch.

Consider the potential
landscape of Fig. \ref{fig1}: a potential well is separated by a ``free''
region by a potential barrier. The well bottom lies below the zero
energy, and is adjusted so that there is only one bound state in the well.
For energies $E>0$, states extend from
the free region to the well region. If the barrier height was infinite
a succession of bound states
would be present in
the well. However, due to the finite width of the barrier separating the
well region from the free region, the excited states become quasi-levels,
which have a finite
lifetime in the well. In the numerical calculations which follow, we
have adjusted the height of the barrier such that only one quasi level is
present among $\sim 100$ ``continuum states''.
By applying a dipole field on this system,
with a frequency which corresponds to the spacing between the ground
state and a quasi level,
the wave function of a particle, initially in the ground state, will
make transitions to the quasi level, and consequently leak out in the
free region if the barrier is not too thick.
The current generated in
the free region then depends on the amplitude and frequency of the
driving field, as well as the characteristics of the barrier.

Typical values for the parameters of the potential of Fig. \ref{fig1}
are $a=4$, $b=2$, $c=200$,
$V=1$, $W=3$. The bound and excited states
wave functions are specified by the connection formulas for
the wave functions at each boundary,
and the corresponding energies of these states are determined numerically.
To determine the energy of the quasi levels, the
integrated density in the well region for
all states with $E>0$ is calculated, and the
level for which this quantity is a maximum is selected. Alternatively,
all matrix elements of the position operator
between the ground and excited states are computed, and the
level for which the probability of transition is a maximum
is selected. In practice,
these two procedures
give the same result.
Once the spacing between the quasi level energy
$E_Q$ and the
ground state energy $E_G$ is known, the driving
frequency $\omega$ is chosen to correspond to an exact
resonance
$\omega=E_Q-E_G$, or
alternatively, one can
include a finite mismatch $\delta\omega$.
At $t=0$, the particle is
taken to be in the ground state, and at each time step,
the integrated density $\rho_i(t)=\int_{a+b}^{a+b+c} d x
{}~\rho(x, t)$
in the free region is computed, as well as the overlap
$|<G|\psi(t)>|^2$ with the ground state.
The time step has to be chosen small compared to the
period of the external field: here, we choose
$\delta t=0.0125 \times (2\pi/\omega)$.

In Fig. \ref{fig2}a, $\rho_i(t)$ is plotted
for field amplitude $\epsilon=0.1$.
The barrier width ($b=2$) and height ($W=3$)
are chosen to be large
enough that the ``escape time'' of the wave function is large compared
to other characteristic times of the problem. Moreover,
the driving frequency has been chosen to be close to the resonance
condition ($\delta\omega/\omega=0.0001$), which is smaller
than the spacing
between ``continuum'' levels ($E>0$), in order to check agreement with
the two level approximation.
The integrated density exhibits steps or plateaux,
which allow to identify the Rabi frequency $\omega_R$. Superposed
to the plateaux structure are small amplitude oscillations which period
corresponds to the driving frequency. As
the simulation is started, transitions to the quasi levels and
neighboring levels
start occurring, but after a period $T_R=2\pi/\omega_R$, the contribution
of the wave function which remained trapped in the well
has returned for the most part in the ground state. This is
illustrated in fig. \ref{fig2}b : indeed, aside from a slow decay
associated with the transparency of the barrier, the overlap with the ground
state $|<G|\psi(t)>|^2$
oscillates with period which matches exactly the plateaux structure.
After the first plateau, $\rho_i(t)$ picks up again as the next oscillation
returns the trapped wave function to the excited states. Upon
doubling/halving
the coupling strength, the period of the oscillations is twice
as small/large,
confirming the fact that the Rabi oscillation frequency scales linearly
with the field amplitude if the resonance condition is met.
The ``measured'' Rabi frequency $\omega_R^m\simeq 0.01$ is in
reasonable agreement
with the two level result $\hbar\omega_R=|<G|x|1>| \epsilon\simeq 0.04$.
To estimate the magnitude of the matrix element $<G|x|1>$ between
the ground state and the quasi level,
the parameters of an infinite well were chosen.

The evolution of the wave function  for the same parameters
is depicted in Figs. \ref{fig3}a--c for times $t=1500$,
$t=3000$, and $t=4500$, where the probability density is
plotted as a function of position. The well and barrier region
lies on the left hand side of the pictures. The Rabi
frequency corresponds roughly to a period $T_R\sim 1500$, and the leakage
current is small, which explains by most of the wave function
remains in the well.
At time $t=1500$ (Fig. \ref{fig3}a), a portion of the wave
function has been transmitted in the free region, as a Rabi
oscillation with the quasi level has been completed.
At $t=3000$ (Fig. \ref{fig3}b), the system has undergone
two Rabi oscillations
and another wave packet escapes from the well, and similarly for
$t=4500$ (Fig. \ref{fig3}c) after three oscillations.
The wave function is therefore emitted by ``bursts''`
out of the well
every time a Rabi oscillation is completed.

In summary, a method has been described, which simulates the quantum
evolution of a
particle in a square barrier/well potential, in the presence of
a dipole field. Using a gauge transformation,
a Hamiltonian which terms commute with each other is obtained,
which in turns allows to write an exact expression for the
evolution operator. Using a generalization of the finite
difference scheme
of (GSS), the time evolution is obtained. This method
is particularly
useful to model activation/tunneling processes for
nano structures
exposed to an external microwave field. As an example,
it is possible to use the two state slow (Rabi) oscillations
of a two level system to build a switching device. The device is composed
of a well region separated from a continuum region by a barrier.
The activation of the ground state to a quasi level using microwave
frequencies allows to generate a current in the free region in
a controllable manner. More details on the characteristics of this
device will be provided elsewhere \cite{Martin}.

\acknowledgements
I thank Gennady Berman for illuminating discussions concerning the
switching device.

\begin{figure}
\caption{Potential landscape considered: a well region is separated by a
potential barrier from the ``free region''. The well depth and barrier height
are adjusted so that there is only one bound state in the well, and there is
only one quasi-level below the barrier}\label{fig1}
\end{figure}
\begin{figure}
\caption{a) Integrated density in the ``free'' region as a function
of time, for $\epsilon=0.1$.
b) Overlap with the ground state as a function of time, with the same
input parameters.}\label{fig2}
\end{figure}
\begin{figure}
\caption{Probability density as a function of position. The well and the
barrier are placed on the left hand side of the pictures. a) $t=1500$
b) $t=3000$ c) $t=4500$ ($t$ is measured in number of time steps
$\delta t=0.0125\times(2\pi/\omega$).}
\label{fig3}
\end{figure}
\end{document}